# Text-Independent Speaker Recognition for Low SNR Environments with Encryption

Aman Chadha
Department of Electronics & Telecommunication
TSEC
Mumbai, India
aman.x64@gmail.com

Divya Jyoti
Department of Electronics & Telecommunication
TSEC
Mumbai, India
dj.rajdev@gmail.com

M. Mani Roja
Department of Electronics & Telecommunication
TSEC
Mumbai, India
maniroja@yahoo.com

## ABSTRACT
Recognition systems are commonly designed to authenticate users at the access control levels of a system. A number of voice recognition methods have been developed using a pitch estimation process which are very vulnerable in low Signal to Noise Ratio (SNR) environments thus, these programs fail to provide the desired level of accuracy and robustness. Also, most text independent speaker recognition programs are incapable of coping with unauthorized attempts to gain access by tampering with the samples or reference database. The proposed text-independent voice recognition system makes use of multilevel cryptography to preserve data integrity while in transit or storage. Encryption and decryption follow a transform based approach layered with pseudorandom noise addition whereas for pitch detection, a modified version of the autocorrelation pitch extraction algorithm is used. The experimental results show that the proposed algorithm can decrypt the signal under test with exponentially reducing Mean Square Error over an increasing range of SNR. Further, it outperforms the conventional algorithms in actual identification tasks even in noisy environments. The recognition rate thus obtained using the proposed method is compared with other conventional methods used for speaker identification.

## 1. INTRODUCTION
Humans have used body characteristics such as face, voice, gait, etc. for thousands of years to recognize each other. Alphonse Bertillon, chief of the criminal identification division of the police department in Paris, developed and then practiced the idea of using a number of body measurements to identify criminals in the mid-19th century. A wide variety of systems require reliable personal recognition schemes to either confirm or determine the identity of an individual requesting their services. The purpose of such schemes is to ensure that the rendered services are accessed only by a legitimate user and thus, disallow unauthorized access [1]. A biometric system is essentially a pattern recognition system that operates by acquiring biometric data from an individual, extracting a feature set from the acquired data, and comparing this feature set against the template set in the database. Thus, biometric systems fall under the ambit of technology designed and used specifically for measuring and analyzing the unique characteristics of a person. Any physiological and/or behavioral characteristic of a person can be used as a biometric feature as long as the following criteria are taken into account [2]:

1) Universality: Each person should have a characteristic which is distinct to the person in question;

2) Distinctiveness: Any two people should be sufficiently different in terms of their characteristics;

3) Permanence: The characteristic should be sufficiently invariant over a reasonable period of time;

4) Collectability: Measuring the characteristic quantitatively should be possible;

5) Performance: This refers to the achievable recognition accuracy and speed, the resources required to achieve the desired performance, as well as the operational and environmental factors that affect performance;

6) Acceptability: It indicates the extent to which people are willing to accept the use of a particular biometric identifier, i.e., the characteristic, in their daily lives;

7) Circumvention: This reflects how easily the system can be bypassed using fraudulent methods.

Even though reliable methods of biometric personal identification like finger-print analysis and retinal or iris scan do exist, however, the validity of forensic fingerprint evidence has recently been challenged by academics, judges and the media. While fingerprint identification was an improvement over earlier systems, the subjective nature of matching, along with the relatively high error rate of has made this forensic practice controversial. As far as iris recognition is concerned, it is very difficult to perform at a distance larger than a few meters and depends on the cooperation of the person. The initial investment in setting these systems is relatively high [2]. In contrast, voice recognition systems have the following advantages over other biometric identification systems:

- Users can enroll themselves over the telephone rather than having them enroll in person to deliver a fingerprint or an iris scan.

- The technology also requires no special data-acquisition system, other than a microphone. In case of signature verification systems, a specialized digital pen tablet acts as the data acquisition system whereas in iris recognition systems, an Iris reader is deployed. Thus, hardware costs are reduced to a minimum.

- The voiceprint generated upon enrolment is characterized by the vocal tract, which is a unique physiological trait. A cold does not affect the vocal tract, so there will be no adverse effect on accuracy levels. Only extreme conditions such as laryngitis can hinder the optimal performance of the system.

- Voice recognition offers relatively low perceived invasiveness as compared to iris recognition, face recognition and signature verification.

Speaker recognition is the process of validating a user's claimed identity using characteristics extracted from their voices. No two individuals sound identical because their vocal tract shapes, larynx sizes and other parts of their voice production organs are different. In addition to these physical differences, each speaker has a distinctive manner of speaking, like the use of a particular accent, rhythm, intonation style, pronunciation pattern, choice of vocabulary etc. [3]. Depending on the context of the application, speaker recognition systems may operate either in verification mode or identification mode.

Speaker identification can be further divided into two branches: open-set identification and closed-set identification. Since it is generally assumed that imposters, i.e., those assuming identity of valid users, are not known to the system, this is referred to as an open-set task. Generally it is assumed the unknown voice must come from a fixed set of known speakers, thus the task is often referred to as closed-set identification. In this paper, we deal with an instance of closed-set speaker identification.

Depending on the algorithm used for the identification, the process can be categorized as text-dependent or text-independent identification. If the text must be the same for enrollment and verification this is called text-dependent recognition. In text-dependent systems, suited for cooperative users, the recognition phrases are known beforehand. For instance, the user can be prompted to read a randomly selected sequence of numbers as illustrated in [7]. In text-independent systems, there are no constraints on the words which the speakers are allowed to use. Thus, the reference and the test utterances may have completely different content. Text-independent systems are most often used for speaker identification as they require very little, if any, cooperation by the speaker. In fact, enrollment may happen without the user's knowledge, as in the case for many forensic applications [3].

Most Biometric Recognition systems are used as security gateways which control access to sensitive information. In such programs if a false negative is triggered, it can be corrected in most cases, by testing again, whereas a false positive may have disastrous consequences from the view point of Data Security and Integrity. Thus, the accuracy rate must be calculated by taking into account both- samples that the program failed to recognize and samples which were identified incorrectly. Further, in text independent speaker recognition systems, an imposter may gain permissions by the following means:

1) A mimicry artist may be employed to imitate the speaker's voice and diction.

2) The speaker's voice may be recorded without his or her consent and knowledge, and this sample may be played to the testing software.

3) The imposter may replace the reference sample of the user in the database by his or her own voice sample.

4) Easily available voice changer software may be used by an imposter to mimic the voice of any reference sample if the database is vulnerable.

All the above scenarios will result in a false positive result due to short comings of a Recognition system based purely on voice. Thus, the paper proposes extensive use of Encryption by means of a private-key which generated from the password selected by the user. This key is then used to seed two levels of Pseudo Random Noise Generators (PRNG) for scrambling the signal sandwiched with Transform-based encryption to increase the robustness of encoding algorithm.

The performance of automatic speech recognizers (ASR) has known to degrade rapidly in the presence of noise and other distortions [6]. Speech recognizers are typically trained on clean speech and typically render inferior performance when used in conditions where speech occurs simultaneously with other unwarranted sound sources, i.e., distortions and disturbances. Some of these unsolicited sources are as below:

- Speech recorded with a microphone or telephone handset is generally vulnerable to environmental noise such as computer hum, car engine, door slams, echoing, keyboard clicks, traffic noise, background noise, which adds to the speech wave [10].

- Reverberation adds delayed versions of the original signal to the recorded signal [4].

- The A/D converter adds its own distortion, and the recording device might interfere with mobile phone radio-waves.

- If the speech is transmitted through a telephone network, it is compressed using lossy techniques which might have added noise into the signal.

To sum up, the speech wave fed to the recognition algorithm is not the same wave that was transmitted from the speaker's lips and nostrils, but it has gone through several transformations degrading its quality [10]. If samples of the corrupting noise source are available before hand, a model for the noise source can additionally be trained and noisy speech may be jointly decoded using trained models of speech and noise [9]. However, in many realistic applications, adequate amounts of noise samples are unavailable before-hand, and hence training of a noise model is not feasible. Fig. 1 illustrates some additive and convolutive noise sources which occur during the process of speaker recognition.

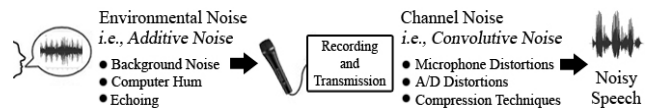

**Fig 1: Error sources during the process of speaker recognition**

The Mel-Frequency Cepstral Coefficients (MFCC) are the most evident example of a feature set that is extensively used in speaker recognition. When MFCC front-end is used in speaker recognition system, one makes an implicit assumption that the human hearing mechanism is the optimal speaker recognizer. However, this has not been confirmed, and in fact opposite results exist [10]. In most speaker recognition systems, MFCC has shown to achieve fairly good performance. Conventionally, MFCC features are extracted from the spectral analysis of 20 to 30 ms long speech frames with an overlap of 10 to 15 ms [17],[18]. The length of the analysis window and the size of overlap are usually fixed for each system. The drawbacks of a fixed length analysis window have been issued by many researchers [18],[23]. This paper presents a modified version of the autocorrelation pitch extraction algorithm robust against noise.

## 2. IDEA OF THE PROPOSED SOLUTION

The primary objective of this paper is to implement a speaker recognition system sustainable to a great extent, against noise, i.e., a system offering superior performance under low SNR conditions. Moreover, for the system to offer high levels of security, a robust multi-level encryption scheme needs to be implemented. The database used for this project consists of voice samples of 50 subjects. 6 voice samples are taken for each person. Out of these, 3 are used for training and the remaining samples are used for testing. Test voice samples with different tones and volume levels are considered for the experiment.

### 2.1 Need for Reference Template Encryption

Encryption is the process of transforming data into scrambled unintelligible cipher text using a key. The role of Encryption is to secure information when it is stored or in transit. However, it is relatively easier to crack single level encryption by brute force or correlation, as compared to a multi-level encryption scheme. Therefore, the program requires a user defined 8-character password to seed two out of three levels of encryption. The password must have a minimum of one capital alphabet, one numeral and one special character, such as @, >, & etc. Fig. 2 shows how, more than $6.6 \times 10^{15}$ permutations are possible. This password is processed further using numeric substitutions in Caesar's Cipher type of encryption and a state seed is generated. Findings in [5] show that a hacker may take as less as 10 minutes to crack each password once a rainbow table has been built, if all passwords are stored internally in a memory hash. Hence, the system is designed to store no password and the state seed generated on the fly is divided into two keys, each of which is used for further encryption and computation.

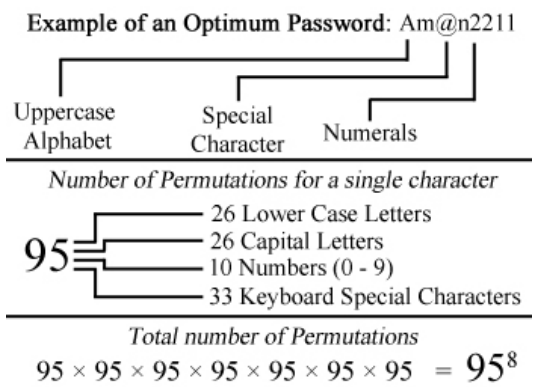

**Fig 2: Permutations for a password**

Random numbers can be generated by a random bit generator which can be defined as a device or algorithm whose output is a sequence of statistically independent and unbiased binary digits. Pseudorandom Number Generator (PRNG) [12] is used to generate random bits dependent on the state seed, such that an adversary cannot judge the next bit by correlating a subset of the random bits generated. This is ensured by virtue of large number of internal states of the generator which in turn means large period of the random bits generated. For implementing the proposed encryption algorithm, PRNG inbuilt in MATLAB 7 is used which has an average period of 235×16 as total number of internal states are 35 words. In short, on an average the random sequence will repeat itself only after 235×16 bits, which is much greater than the length of the samples required for testing.

PRNG works by taking a state seed s as input and generating the output sequence of random values as f(s), f(s+1), f(s+2), … Here, f is a one way function which can be defined as an algorithm whose output is a sequence of statistically independent and unbiased binary digits [12]. Based on the properties of this function f, some output values say, f(s+i), must be discarded to eliminate correlation between subsequent random bits. This approach is actively followed by standardized one-way functions, for e.g., a cryptographic hash function such as SHA-1 or a block cipher such as DES (x7.4). When the sequence so generated is superimposed with the reference or test voice sample, scrambling takes place and signal is rendered nearly indecipherable. Yet, the signal is still in time domain and a person may make a correct guess of the scrambling key. Thus, the second type of encryption used is Transform based encryption.

A transform based encryption changes the domain of the data from say time domain to frequency domain or complex plot etc. It means that the new data which is acquired has no meaning in its previous domain, in this case-time domain. A discrete cosine transform (DCT) based scheme is used for further scrambling as [11] illustrates the superiority of DCT over four other discrete transform based encryption techniques for analog speech, when compared with respect to a novel cryptanalytic attack. DCT is a well-known transform that decomposes a signal into its frequency components and it un-correlates the sequence of input samples, i.e. DCT coefficients give the frequency domain equivalent of speech data [11]. In fact, for large databases this step may be used to compress the voice signal at the cost of system accuracy. A final layer of pseudorandom noise using the second part of the state key generated at run-time is superimposed on the noisy signal DCT coefficients and source side encryption is complete. The sample so obtained is amplified and assumed to be transmitted over an AWGN channel with known Signal to Noise Ratio (SNR). At the testing site, this received sample is decrypted in inverse order of encryption and matched with test sample for Speaker Recognition.

### 2.2 Speaker Recognition Process

Fig. 3 shows the general structure of the speaker recognition system.

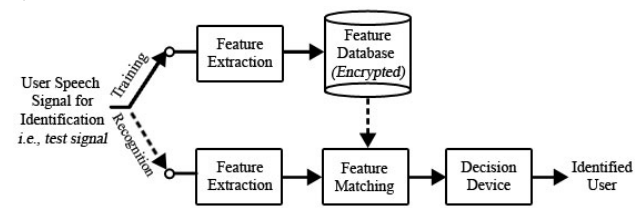

**Fig 3: General architecture of a speaker recognition system**

This system operates in two modes: training and recognition. In the training mode a new speaker (with a known identity) is enrolled into the database, while in the recognition mode an unknown speaker gives a speech input signal and the system tries to identify the speaker.

1) Feature Extraction: The feature extractor, i.e, the front-end, is the first component in an ASR system. Feature extraction transforms the raw speech signal into a compact but effective representation that is more stable and discriminative than the original signal.

2) Database: A collection of voice samples has been recorded for evaluation of the proposed system. The recordings were converted to WAV format to facilitate easy analysis and operations using MATLAB 7. The WAV files were subjected to a multi-level encryption scheme to foster maximum security.

3) Speaker Modeling: The training phase uses the acoustic vectors extracted from each segment of the signal to create a speaker model which will be stored in a database.

4) Pattern matching and decision: The Pattern matching strategy takes all the matching scores from the user pattern to each of the stored reference patterns into account and searches for the "closest" possible match and thus makes a decision.

The steps involved in the entire process can be summarized as follows:

1) The locations of the samples of different users are stored upon encryption them using the proposed multi-level encryption scheme.

2) Various features and parameters of the voice samples such as mean, variance, standard deviation, pitch etc. are calculated.

3) The voice sample of a test speaker is recorded.

4) The Euclidean distance between the features of this sample and the samples previously stored in the database is calculated.

5) The Euclidean distances are then arranged in an ascending order, with the first Euclidean distance being the minimum.

6) The sample corresponding to the first Euclidean distance is the sample having the highest resemblance to the sample under test.

## 3. IMPLEMENTATION STEPS

Pitch, i.e., fundamental frequency, is an important parameter of speech signals which is used in speech analysis, synthesis and recognition. Fundamental frequency (F0) as an acoustic correlate is strongly related to prosodic information of stress and intonation. For speech recognition applications, pitch extraction (fundamental frequency estimation) provides the basis for voiced/unvoiced classification decision. However, before pitch extraction and subsequent matching takes place, the Reference voice sample must be encrypted, transmitted over AWGN channel and decrypted at testing site. Fig. 4 shows the data flow at various steps.

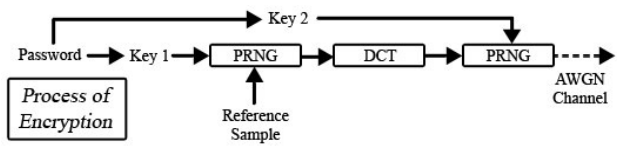

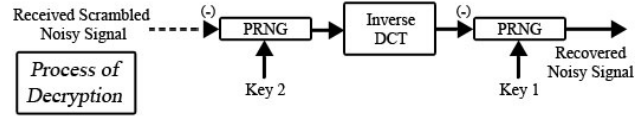

**Fig 4: Scheme for encryption and decryption**

The steps involved in the process of generation of the two State Keys can be summarized as follows:

1) Read an 8 character long password from user.

2) Convert to ASCII equivalent to form an array say 'x'.

3) Apply Caesar's cipher with a shift of 4, to each ASCII element to form a new array say 'y'.

4) Concatenate all elements of y to form a single integer say 'z'.

5) Calculate length of 'z'.

6) If even break into two equal halves two generate two equal length keys – Key1 and Key2.

7) Else break asymmetrically, the longer key is Key1 and shorted key is Key2.

Sample Password: Djyot!24

ASCII Array x: [68 106 121 64 116 105 50 52]

Cipher Text Array y: [72 110 125 69 120 109 54 56]

Concatenated Integer z: 72110125691201095456

Key1: 7211012569         Key2: 1201095456

**Fig 5: An example illustrating generation of keys**

The steps involved in the process of encryption can be summarized as follows:

1) Level 1: Key1 is passed as seed to PRNG and superimpose the output sequence on reference voice sample to generate a noisy signal say 'x'.

2) Level 2: Perform DCT on x and save the coefficients in an array say 'y'.

3) Level 3: Key2 is passed as seed to PRNG and superimpose the random sequence obtained on y and save the resultant as an array say 'z'.

4) z is the signal which is to be transmitted through AWGN channels with known SNR. Thus, it is subjected to various SNR levels and the resulting signal is normalized and then decrypted as follows.

The steps involved in the process of decryption can be summarized as follows:

1) Level 3: Key2 is passed as seed to PRNG and algebraically subtract the random sequence so obtained from the received signal, save the resultant as say 'x'.

2) Level 2: Take Inverse DCT of x and save the noisy signal so obtained as 'y'.

3) Level 1: Key1 is passed as seed to PRNG and algebraically subtract the random sequence so obtained from x, save the resultant as say, 'z'.

4) The file 'z', recovered after the above steps is the final decrypted version of reference signal, this is sent for pitch extraction along with the test signal and comparative matching takes place.

Fig. 6 and Fig. 7 show the encryption and decryption plots of the sample under test respectively.

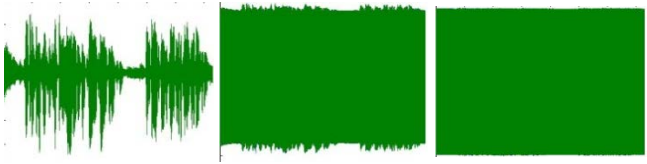

**Fig 6: Original signal (Left), signal after Level 2 (Middle) and signal after Level 2 and AWGN Insertion (Right)**

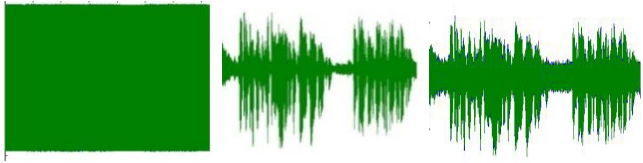

**Fig 7: Received signal (Left), recovered file with high SNR (Middle) and recovered file with low SNR (Right)**

Pitch extraction, also known as, fundamental frequency estimation, plays a vital role in speech processing and has numerous applications in speech related areas. Therefore, several methods to extract the pitch of speech signals have been proposed by researchers. However, such methods are known to be very vulnerable in noisy environments, hence, performance improvement in noisy environments is still desired. For example, this is particularly true in speech enhancement systems, because in such systems the accuracy of pitch extraction is directly related with the quality of speech after the operations of enhancement. Also, speech communication systems often transmit pitch information. To do this, we have to extract the pitch of speech signals in practical noisy environments. Unfortunately, a reliable and accurate method for pitch extraction in noisy environments is still a subject of scientific investigation.

Generally, pitch detection algorithms (PDA) use short-term analysis techniques [16]. For every frame $x_m$ we get a score $f(T|x_m)$ which is a function of the candidate pitch period $T$. Such algorithms, in general, offer a rough estimation of the pitch by maximizing the following equation:

$$T_m = \underset{T}{\arg\max}\, f(T \mid x_m) \qquad (1)$$

A commonly used method to estimate pitch is based on detecting the highest value of the autocorrelation function in the region of interest. The correlation between two waveforms is a measure of their similarity. The waveforms are compared at different time intervals, and their similarity is calculated at each interval. The result of a correlation is a measure of similarity as a function of time lag between the beginnings of the two waveforms. One would expect exact similarity at a time lag of zero, with increasing dissimilarity as the time lag increases. The mathematical definition of the autocorrelation function $R_{xx}(\tau)$ is shown in (2), for an infinite discrete function $x[n]$, and (3) shows the mathematical definition of the autocorrelation $R_{xx}(\tau)$ of a finite discrete function $x'[n]$ of size $N$.

$$R_{xx}(\tau) = \sum_{n=-\infty}^{\infty} x[n]x[n+\tau] \qquad (2)$$

$$R_{xx}'(\tau) = \sum_{n=0}^{N-1-\tau} x'[n]x'[n+\tau] \qquad (3)$$

where, $x[n]$ is the speech signal;

$\tau$ is the lag number;

$n$ is the time for a discrete signal.

Correlation based processing is known to be comparatively robust against noise and may be one which provides the best performance in noisy environments [19],[20],[22]. The autocorrelation function of a signal is actually a non-invertible transformation of the signal that is useful for displaying structure in the waveform. Hence for pitch detection applications, if we assume $x(n) = x(n + P)$ for all $n$, i.e., $x(n)$ is periodic with period $P$, then it is easily shown that:

$$R_{xx}(\tau) = R_{xx}(\tau + P) \qquad (4)$$

Equation (4) basically indicates that the autocorrelation function is also periodic with the same period. Conversely, periodicity in the autocorrelation function indicates periodicity in the original signal.

Speech being a non-stationary signal, the concept of a long-time autocorrelation measurement as defined in (3) cannot be easily extrapolated for such signals [16]. Thus, it is mandatory to define a short-time autocorrelation function, which operates on short segments of the signal as:

$$R_{xx}(\tau) = \frac{1}{N}\sum_{n=0}^{N'-1}\left[x(n+l)w(n)\right]\left[x(n+l+\tau)w(n+\tau)\right] \qquad (5)$$

where, $0 \leq \tau \leq M_0$;

$w(n)$ is an appropriate window for analysis;

$N$ is the section length being analyzed;

$N'$ is the number of signal samples used in the computation of $R_{xx}(\tau)$;

$M_0$ is the number of autocorrelation points to be computed;

$\tau$ is the lag number;

$l$ is the index of the starting sample of the frame.

For applications involving pitch estimation, $N'$ is generally set to the value given by the following equation:

$$N' = N - m \qquad (6)$$

This is done so that only the N samples present in the analysis frame, i.e., $x(l), x(l + 1), \ldots, x(l + N - 1)$ are used in the autocorrelation computation. Values of 200 and 300 have generally been used for $M_0$ and $N$ respectively corresponding to a maximum pitch period of 20 ms (200 samples at a 10 kHz sampling rate) and a 30 ms analysis frame size.

Correlation based processing also includes the average magnitude difference function (AMDF) method [15],[16],[18]. The AMDF PDA is chosen in our study is because it has relatively low computational cost and is easy to implement. Several types of noise such as babble, car, and street noises with 5 dB, 10 dB, 15 dB and 20 dB SNR are used to evaluate the performance of the proposed system. The noise sources are taken from the NOISEX-92 database [21], from Carnegie Mellon University, a collection of different noise waveforms which can be used to generate speech waveforms in various noise conditions and with different signal to noise ratio (SNR) values. The AMDF is [13] essentially a variation of autocorrelation function analysis where, instead of correlating the input speech at various delays (where multiplications and summations are formed at each value), a difference signal is formed between the delayed speech and the original, and at each delay value the absolute magnitude is taken [14]. In contrast with the autocorrelation or cross-correlation function, the AMDF calculations require no multiplications, a much sought-after property for real-time applications. Therefore, for the purpose of emphasizing the true peak produced by the autocorrelation, i.e., for measuring the periodicity of voiced speech, we propose an autocorrelation function (AMDF). The AMDF is defined by the following equation:

$$AMDF(\tau) = \frac{1}{N} \sum_{n=0}^{N-1} |x[n] - x[n-\tau]| \quad (7)$$

where, $x(n)$ are the samples of input speech;

$x(n - \tau)$ are the samples obtained by introducing a delay of $\tau$ seconds.

Equation (7) indicates the characteristic of the AMDF that when x[n] is similar with x[n - τ], AMDF(τ) yields a small value. A difference signal is thus formed by delaying the input speech various amounts, subtracting the delayed waveform from the original and summing the magnitude of the differences between sample values. For zero delay, the difference signal is always zero and is particularly small at delays corresponding to the pitch period of a voiced sound having a quasi-periodic structure [16].

For each value of delay, computation is made over an integrating window of *N* samples. To generate the entire range of delays, the window is "cross differenced" with the full analysis interval. An advantage of this method is that the relative sizes of the nulls tend to remain constant as a function of delay, which is mainly because there is always full overlap of data between the two segments being cross differenced. In extractors of this type, the limiting factor on accuracy is the inability to completely separate the fine structure from the effects of the spectral envelope. For this reason, decision logic and prior knowledge of voicing are used along with the function itself to help make the pitch decision more reliable.

Let us assume that *x(n)* is a noisy speech signal composed of the actual speech content and the Additive White Gaussian Noise (AWGN). *x(n)* is given by the following equation:

$$x(n) = s(n) + w(n) \quad (8)$$

where, *s(n)* is a clean speech signal;

*w(n)* is the AWGN.

The autocorrelation function $R_{xx}(\tau)$ for this particular case, as demonstrated in [15], is given by:

$$= \frac{1}{N} \sum_{n=0}^{N-1} (s[n] + w[n]) \cdot (s[n+\tau] + w[n+\tau])$$
$$= \frac{1}{N} \sum_{n=0}^{N-1} (s[n]s[n+\tau] + s[n]w[n+\tau] + w[n]s[n+\tau] + w[n]w[n+\tau])$$
$$= R_{ss}(\tau) + 2R_{sw}(\tau) + R_{ww}(\tau)$$

where, $R_{ss}(\tau)$ is the autocorrelation function of *s[n]*;

$R_{sw}(\tau)$ is the crosscorrelation function of *s[n]* and *w[n]*;

$R_{ww}(\tau)$ is the autocorrelation function of *w[n]*.

In [15], the case for large values of N has been described. If the speech signal shows no correlation with the AWGN, then $R_{sw}(\tau)$ does not exist, i.e., it yields a zero value. The following equation exists for this case:

$$R_{xx}(\tau) = R_{ss}(\tau) + R_{ww}(\tau) \quad \text{...if } \tau = 0 \quad (8)$$

Also, if *w[n]* is uncorrelated, then $R_{ww}(\tau)$ yields a zero value except for *τ = 0*. If this is the case, the following relation holds true:

$$R_{xx}(\tau) = R_{ss}(\tau) \quad \text{...if } \tau \neq 0 \quad (9)$$

Based on the above mentioned properties, the autocorrelation function provides robust performance against noise. When the characteristics of the AMDF are plotted, it is found to yield a notch, while the autocorrelation function yields a peak. Thus, the characteristics of the AMDF are found to bear similarity with that of the autocorrelation function. However, both functions have the same periodicity. Pitch of the segmented speech is estimated by searching the peak of the resultant function obtained on coupling the autocorrelation function with the AMDF. However, upon deploying the resultant function directly, we observe that the accuracy of pitch extraction is compromised. Therefore, the system uses interpolation based on 3 points around the detected peak [15]. It is known that such interpolation on the autocorrelation function is useful for improving the accuracy of pitch extraction. Lagrange's method is used to perform the interpolation operation. The frequency band selected for searching the pitch peak is from 50 Hz to 400 Hz as it corresponds to the region of the fundamental frequencies of most men and women. The technique of removing the formant structure for reliable pitch detection by center clipping demonstrated by M. Sondhi [8] while still retaining the pitch period information was implemented to reduce the effects of the formant structure on the detailed shape of the short-time autocorrelation function.

## 4. RESULTS

Upon performing Decryption on the signals that were subjected to AWGN, the Mean Square Error (MSE) values thus obtained, are tabulated against SNR as follows:

**Table 1. MSE values for corresponding SNR values**

| SNR in dB | Mean Square Error |
|---|---|
| 16 | $2.83 \times 10^{-2}$ |
| 17 | $1.82 \times 10^{-2}$ |
| 18 | $2.02 \times 10^{-3}$ |
| 19 | $4.11 \times 10^{-5}$ |
| 20 | $7.32 \times 10^{-7}$ |

Fig. 8 shows the plot of the MSE values against the SNR values.

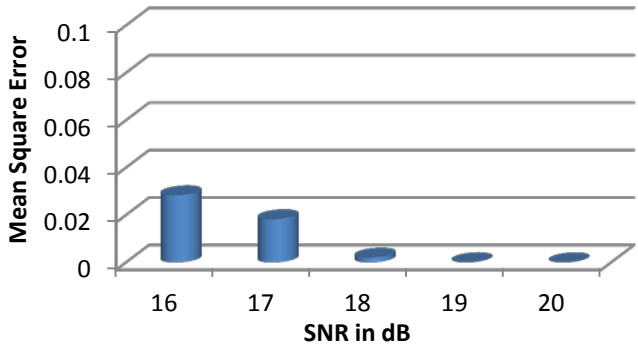

**Fig 8: Plot of MSE against corresponding SNR**

To investigate the accuracy of the modified pitch extraction method, various experiments were conducted to compare the efficiency of the algorithm with four standard conventional speaker identification methods namely, Statistical Methods (Mean, Moment and Variance), Linear Predictive Coding (LPC), Zero Crossing and Fast Fourier Transform (FFT).

**Table 2. Results obtained using the proposed modified autocorrelation method and comparison with other methods**

| Algorithm | Accuracy |
|---|---|
| Pitch Extraction | 92.39 |
| Mean, Moment and Variance | 74.87 |
| Linear Predictive Coding | 72.42 |
| Zero Crossing | 62.35 |
| Fast Fourier Transform | 55.49 |

Fig. 9 shows the plot of accuracy rates of various algorithms.

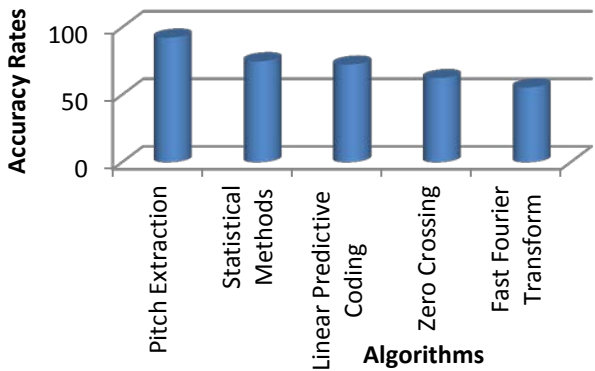

**Fig 9: Plot of various algorithms against their accuracy rates**

## 5. FUTURE SCOPE
The system can be implemented for real time applications if the database can be standardized online, this will eliminate the need of training samples for the system. Also, the concept of voice recognition using autocorrelation can be tried for a larger database. As the size of the database increases, encryption algorithm could be further strengthened by using longer keys and increasing the period of pseudorandom noise sequences which shall make decryption by brute force near impossible and also reduce MSE to a minimum.

## 6. CONCLUSION
The primary objective of the paper was to implement a robust and secure voice recognition system using minimum resources offering optimum performance in noisy environments. We have implemented this system using three levels of encryption for data security and autocorrelation based approach to find the pitch of the sample. The resulting system was found to reduce significantly the amount of test data or features to be extracted. By virtue of DCT based scrambling, the system is highly immune to cryptanalytic attacks that target the redundancy of speech. The robustness of the system in adverse conditions such as noisy or channel distorted environments was verified by conducting closed set text-independent speaker identification experiments, and results pointed to improved performance in adverse SNR environments. We conclude that the proposed algorithm is equipped with means to ensure that data security is not compromised at any stage of computation and at the same time high accuracy rate of the pitch detection algorithm makes it very powerful in both clean and noisy environments.

## AUTHOR BIOGRAPHIES

**Aman Chadha** (M'2008) was born in Mumbai (M.H.) in India on November 22, 1990. He is currently pursuing his undergraduate studies in the Electronics and Telecommunication Engineering discipline at Thadomal Shahani Engineering College, Mumbai. His special fields of interest include Image Processing, Computer Vision (particularly, Pattern Recognition) and Embedded Systems. He has 5 papers in International Conferences and Journals to his credit. He is a member of IETE, IACSIT and ISTE.

**Divya Jyoti** (M'2008) was born in Bhopal (M.P.) in India on April 24, 1990. She is currently pursuing her undergraduate studies in the Electronics and Telecommunication Engineering discipline at Thadomal Shahani Engineering College, Mumbai. Her fields of interest include Image Processing, and Human-Computer Interaction. She has 4 papers in International Conferences and Journals to her credit.

**M. Mani Roja** (M'1990) was born in Tirunelveli (T.N.) in India on June 19, 1969. She has received B.E. in Electronics & Communication Engineering from GCE Tirunelveli, Madurai Kamraj University in 1990, and M.E. in Electronics from Mumbai University in 2002. Her employment experience includes 21 years as an educationist at Thadomal Shahani Engineering College (TSEC), Mumbai University. She holds the post of an Associate Professor in TSEC. Her special fields of interest include Image Processing and Data Encryption. She has over 20 papers in National / International Conferences and Journals to her credit. She is a member of IETE, ISTE, IACSIT and ACM.